\documentstyle[12pt]{article}
\def \z{&\hspace*{-8pt}}
\newcommand{\Ls}[2]{{\mbox{Ls}}_{#1}\!\left(#2\right)}

\begin{document}

\vspace{15mm}

\begin{center}

{\large \bf Analytical results for the four-loop RG functions
    in the $2D$ non-linear $O(n)$ $\sigma$-model on the lattice}

\vspace{15mm}

{\large

\vspace{5mm}

Oleg~Veretin%
\footnote{~E-mail: veretin@mail.desy.de}
}


{\it ~II. Institut f\"ur Theoretische Physik,\\
      Universit\"at Hamburg, Hamburg, Germany}

\vspace{15mm}

\end{center}

\begin{abstract}
We recalculate four-loop renormalization group functions 
in 2-dimensional non-linear $O(n)$
$\sigma$-model using coordinate-space method. The high accuracy
of calculation allows us to find the analytical form of $\beta$-
and $\gamma$-function (anomaluos dimension).
\end{abstract}


\vspace*{10pt}

PACS number(s): 11.15.Ha, 11.10.Gh,12.20.Ds

\thispagestyle{empty}
\setcounter{page}0
\newpage

\section{Introduction}

  Non-linear $\sigma$-models have been the objects of the intensive studies
for many years. The particular case of these models, 
considered in this paper, is the 2-dimensional 
non-linear $O(n)$ $\sigma$-model.
This model is known to be asymptotically free and can be applied, e.g.
to the study of ferromagnetic systems. It can also
serve as a toy model for the strong interactions in particle physics.

  In calculations of physically interesting characteristics it is important
to know the $\beta$-function and anomalous dimension $\gamma$. 
The know of them allows, in particular, to predict the correlation 
length $\xi$ and the spin susceptibility $\chi$.
In the regime of weak coupling $\beta$- and $\gamma$-function can be
evaluated as perturbative series in the coupling constant.
In order to study the whole range of the coupling constant one has 
to appeal to the lattice simulations. Due to the asymptotical freedom
this model is especially suitable for such study.
For the precise comparision of Monte Carlo data with perturbative
expansions, higher loop calculations within the lattice regularization 
are required.
Such calculation to two-loops have been done analytically in \cite{lat1} 
and then pushed forward to four-loops in \cite{lat2} numerically and 
checked in \cite{ShinRG}.
At the same time the analogous results at the four-loop order 
in the continuum limit are known analytically \cite{MSfunctions}.

  The goal of this work is to find the analytical expressions for the 
renormalization group (RG)
coefficients to the four-loop order in the lattice perturbation theory.
In order to do this we use the methods proposed in the continuum field
theory for the evaluation of the multiloop integrals. Diagrams
on the lattice, as well as in the continuum limit, are related to each
other algebraically. Such relations arise due to the integration by part 
method \cite{IBP}, which leads, in general, to the reduction of the number
of independent integrals. However the realization of this algorithm on the 
lattice  already at the three-loop level is quite difficult task. 

  In Section 2 we give the difinitions and discuss the method.
In Section 3 our results are presented and in Appendix A we
give all integrals from \cite{lat2} separately.

\section{Definitions}

  The action of the non-linear $O(n)$ $\sigma$-model is usually 
written in the form  
\begin{equation}
S = \frac{1}{2f_0}\int {\rm d}^2 x 
   \Bigl( \partial_\mu q(x)\, \partial_\mu q(x) \Bigr),
\end{equation}
where $q_i(x)$ is an $n$-component real vector field of unit length
and $f_0$ be the bare coupling constant. 
In the lattice formulation the derivatives
are, as usually, understood as finite differences.

  The perturbative expansions of
the $\hat{\beta}$- and $\hat{\gamma}$- functions can be written
as follows%
\footnote{Our coefficients $\hat{b}^{(L)}$ and $\hat{c}^{(L)}$ are
          difined slightly different than these in \cite{lat2}.}
\begin{eqnarray}
\label{betaL}
\hat\beta(f) \z=\z -a\frac{d}{da}f_0 = - 2\pi (n-2) \sum\limits_{L=1}
     \hat b^{(L)} \left(\frac{f_0}{2\pi}\right)^{L+1}, \\
\label{gammaL}
\hat\gamma(f) \z=\z a\frac{d}{da}\ln Z =  2\pi (n-1) \sum\limits_{L=1}
     \hat c^{(L)} \left(\frac{f_0}{2\pi}\right)^{L+1},
\end{eqnarray}
where $a$ is the lattice spacing and $Z$ is the renormalization 
constant of the field. Prefactors $(n-2)$ and $(n-1)$ in the above formulae
always factorize and we take them in front of the expressions.

  Coefficients $\hat{b}^{(L)}$ and $\hat{c}^{(L)}$ can be computed 
using technique of Feynman diagrams. 
Generally, Feynman diagrams on the lattice are more
difficult to evaluate than the ones in the continuum field theory.
Therefore the analytical results in the lattice are known only
to two loops \cite{lat1}, while analogous quantities in the 
continuum theory are known to four loops \cite{MSfunctions}. 
The RG coefficients
were computed on the lattice numerically to four loops \cite{lat2},
where they were expressed in terms 12 different integrals.
The evaluation of this integrals has been
repeated in \cite{ShinRG} to somewhat better accuracy (about $\sim 10^{-9}$) 
and the wrong notation of \cite{lat2} was clarified in \cite{Err}.

  It is known that between different
Feynman diagrams there are many algebraic relations, which can be
obtained by partial integration \cite{IBP}. This explains
the fact that a big amount of different integrals could be expressed
as linear combinations of few constants 
(irrationalities) with rational coefficients.
Moreover there were proposed some rules how to predict the constants
that occur in higher loop calculations \cite{Broadhurst,Kalmykov1}.
The interesteng question arises: which constants appear in the
lattice diagrams calculation? We make a conjecture that they are
the same as in the continuum case, proposed in \cite{Kalmykov1}.
To test this conjecture the so-called PSLQ test \cite{PSLQ} has been used.

  Let as briefly describe this approach. Suppose that we have
some irrational numbers $\eta_1,\dots,\,\eta_n$ given to
some a certain precision with $d$ decimal digits.
We say that they obey an integer relation with norm bound $N$
if $\eta_1,\dots,\,\eta_n$ are linear dependent with integer
coefficients. Precisely, there are exist {\it integer} numbers 
$c_1,\dots,\,c_n$ such that
\begin{equation}
\label{intrel}
\left| c_1\eta_1 + \dots + c_n\eta_n  \right| < \epsilon,
\quad \mbox{provided that} \qquad  {\rm max} |c_i| < N,
\end{equation}
where $\epsilon>0$ is some small number of the order $10^{-d}$ and $N$ is norm bound.

  Given accuracy $d$, ''detection threshold'' $\epsilon$ and
norm bound $N$, the PSLQ test allows to find out whether 
relation (\ref{intrel}) exists or not (see details in \cite{PSLQ}).
This approach has been applied in several calculations 
(see e.g. \cite{pslqapp}).

  The crucial point is the knowlege of the basis elements $\eta_j$.
We suppose, naturally, that the basis for latice integrals
under consideration is the same as for those in continuum field theory
for a single scale diagrams. The reason for that is that the finite
part of diagrams contains the same class of functions, independent
on wich kind of regularization has been used.
It was suggested in \cite{Broadhurst} 
that the basis elements form an algebra: i.e. if 
$\eta_1$ and $\eta_2$ belong to the basis then the product $\eta_1\eta_2$ does either.
Thus some ''higher'' elements (but not all of them) 
are constructed from ''lower'' ones
by forming all possible products of the latters. In addition,
the integral and the basic elements can be ordered by their 
''weights'' (see details in \cite{Broadhurst,Kalmykov1}),
which are determined by the number of loops but not, by the topology of a diagram
(for several single scale diagrams it has been tested in \cite{binomial}).

  Thus we come to the following basis elements
\begin{eqnarray}
\label{even}
&& \pi,\, \log2, \nonumber\\
&& \pi^2,\, \pi\log2,\, \log^2 2,\, G,    \\
&& \pi^3,\, \pi^2\log2,\, \pi\log^2 2,\, \log^3 2,\, 
   G\pi,\, G\log2,\, \zeta_3,\,  {\rm Ls}_3(\pi/2) \nonumber
\end{eqnarray}
and
\begin{eqnarray}
\label{odd}
&& \frac{\pi}{\sqrt3},\, \log3, \nonumber\\
&& \pi^2,\, \frac{\pi}{\sqrt3}\log3,\, \log^2 3,\, 
   \frac{{\rm Ls}_2(\pi/3)}{\sqrt3}, \\
&& \frac{\pi^3}{\sqrt3},\, \pi^2\log3,\, \frac{\pi}{\sqrt3}\log^2 3,\, 
   \log^3 3,\, 
   \pi{\rm Ls}_2(\pi/3),\, \log3 \frac{{\rm Ls}_2(\pi/3)}{\sqrt3},\, 
   \zeta_3,\,  \frac{{\rm Ls}_3(2\pi/3)}{\sqrt3},  \nonumber
\end{eqnarray}
where $\zeta_k=\zeta(k)$ is Riemann $\zeta$-function, $G=0.915965594177219015\dots$
is the Catalan constant
and the constant ${\rm Ls}_2(\pi/3)=1.014941606409653625\dots$ is defined 
through the so-called log-sine integral \cite{Lewin}
$$
{\rm Ls}_k(\theta)=\int\nolimits_0^\theta 
     \log^{k-1}\left( 2\sin\frac{\theta'}{2} \right)\,d\theta'.
$$
In Eqs. (\ref{even}) and (\ref{odd}) the first, second and third
lines correspond to weights 1,\, 2 and 3 respectively.
The elements of higher weights would correspond to higher loop
integrals and not do appear here.

\section{Results and discussion}

  We applied the ideas explained above to the lattice integrals
presented in \cite{lat2}.
The integrals were computed to accuracy better than $10^{-40}$
using the coordinate-space method proposed in \cite{ShinX}.
The most problematic integrals $V_3$ and $V_6$ were computed even 
to higher accuracy.
The analysis established that these integrals can be
expressed within bases (\ref{even}) and (\ref{odd}) plus one more constant,
introduced below. From 28 elements of (\ref{even}) and (\ref{odd}) only
5 do contribute.
Namely, we were able to express all integrals evaluated numerically 
in \cite{lat2,ShinRG} in terms of the following six irrational constants
\begin{equation}
\label{constants}
  \pi,\quad \pi^2,\quad \zeta_3,\quad G,\quad \frac{\Ls{2}{\pi/3}}{\sqrt{3}},\quad
   \quad\mbox{and}\quad (2\pi)^3 K,
\end{equation}
where integral $K$ is the same three-loop bubble as in \cite{lat2,ShinRG}.

Among these integrals only for $K$ we did not find
a relation to the bases (\ref{even}) and (\ref{odd}). 
Therefore we include it
as an independent constant. However it is not excluded that $(2\pi)^3K$
can be rewritten as a linear combination of elements (\ref{even}) and (\ref{odd}) 
and the possible reason for our misfinding is the lack of the accuracy
for the numerical value of this integral.

  For the last constant $K$ we give numerical result accurate to $10^{-37}$
\begin{equation}
(2\pi)^3 K = 23.7849506237378578142256363314563137344(1).
\end{equation}

  Coefficients $\hat b^{(L)}$ of beta function (\ref{betaL})
now read
\begin{eqnarray}
\hat{b}^{(1)} \z=\z 1,\\
\hat{b}^{(2)} \z=\z 1,\\
\hat{b}^{(3)} \z=\z \frac{n-7}{24}\pi^2 + \frac12 \pi - \frac{n-4}{2},\\
\hat{b}^{(4)} \z=\z 
         - \frac{28n^2-66n-38}{12}\zeta_3
         - \frac{(n-2)(n+1)}{8}(2\pi)^3 K
         + \frac{3n-1}{12}\pi^3 
\nonumber\\
\z\z
         - 10(n-2)\pi\frac{\Ls{2}{\pi/3}}{\sqrt{3}}
         + 20(n-2)\pi G 
         + \frac{6n^2-26n-1}{12}\pi^2
\nonumber\\
\z\z
         - 2(n-2)(n+20)\frac{\Ls{2}{\pi/3}}{\sqrt{3}}
         - 4(n-2) G
         - \frac{5n-12}{2}\pi
         + \frac{2n^2-3n-1}{2}.
\end{eqnarray}

  For the anomalous dimension (\ref{gammaL}) we have
\begin{eqnarray}
\hat{c}^{(1)} \z=\z 1,\\
\hat{c}^{(2)} \z=\z \frac12 \pi,\\
\hat{c}^{(3)} \z=\z \frac{n+9}{24}\pi^2 - \frac{n-2}{2},\\
\hat{c}^{(4)} \z=\z 
           \frac{(n-2)(127n-121)}{24}\zeta_3
         + \frac{(n-2)(n+1)}{16}(2\pi)^3 K
         - \frac{4n-11}{24}\pi^3
\nonumber\\
\z\z
         + 5(n-2)\pi\frac{\Ls{2}{\pi/3}}{\sqrt{3}}
         - 10(n-2)\pi G
         - \frac{3n^3-11n+2}{6}\pi^2
\nonumber\\
\z\z
         + (n-2)(7n+8)\frac{\Ls{2}{\pi/3}}{\sqrt{3}}
         + (2n-4)G
         + \frac{13(n-2)}{4}\pi
\nonumber\\
\z\z
         - \frac{(n-2)(10n-21)}{2}.
\end{eqnarray}

  In conclusion, we expressed RG functions within the lattice
regularization in terms of six irrational constants given by (\ref{constants}).
The algebraic structure of the above results suggests that
there should exist a method of algebraic reduction of diagrams
to a set of a few master intergrals. As it is mentioned in the
begining of the paper such method exists in continuum field theory
and is based on the integration by parts \cite{IBP} in the momentum space. 
On the lattice however reduction algorithms are not so obvious.
In the simplest case of vacuum one-loop bubble diagrams 
algebraic method was discussed in \cite{bubble}. In more complicated
cases only few investigations has been done in this directions
(see e.g. \cite{Melnikov}).
The development of algebraic methods is desirable and they
could be very useful tools for higher loop computations on the
lattice.

\noindent
{\bf Acknowledgements}
This work was supported in part by the German Federal Ministry for Education and
Research BMBF through Grant No. 05H12GUE and by the Helmholtz Association HGF
through Grant No. Ha 101.


\appendix

\section{Integrals}

  In this appendix we present separately our analytical
results for the integrals that enter RG functions.
They are given in \cite{ShinX} and \cite{ShinRG} numerically.
So our results for these integrals read
\begin{eqnarray}
\z\z (2\pi)^2 G_1 = \frac12\zeta_2 + 1,\\
\z\z (2\pi)^2 R = \frac{\Ls{2}{\pi/3}}{\sqrt{3}},\\
\z\z (2\pi)^3 J = -24\zeta_2\pi+96\zeta_2,\\
\z\z (2\pi)^3 L_1 =
          -\frac72\zeta_3 + 3\zeta_2,\\
\z\z (2\pi)^3 V_1 =
                  \frac72\zeta_3 ,\\
\z\z (2\pi)^3 V_2 = \frac{14}{3}\zeta_3 - 4\zeta_2
            + 8\frac{\Ls{2}{\pi/3}}{\sqrt{3}} - 4,\\
\z\z (2\pi)^3 V_3 =  \frac{56}{3}\zeta_3 
            - 16\zeta_2 + 24\frac{\Ls{2}{\pi/3}}{\sqrt{3}} -16 + (2\pi)^3 K, \\
\z\z (2\pi)^3 V_4 = - \frac{13}{24}\zeta_3,\\
\z\z (2\pi)^3 V_5 = \frac{19}{2}\zeta_3 - 3\pi\zeta_2 + 4\zeta_2,\\
\z\z (2\pi)^3 V_6 =  \frac{14}{3}\zeta_3 
                       - 8\zeta_2 + \frac{1}{2}(2\pi)^3 K, \\
\z\z (2\pi)^2 W_1 = - \frac12 \frac{\Ls{2}{\pi/3}}{\sqrt{3}}, \\
\z\z (2\pi)^3 \hat W_2 = \frac12 \zeta_3
         + \frac32 \pi \frac{\Ls{2}{\pi/3}}{\sqrt{3}}
         - \frac52 \pi G
         + \frac12 \zeta_2
         + \frac{11}{2} \frac{\Ls{2}{\pi/3}}{\sqrt{3}}
         + \frac12 G
         - \frac12.
\end{eqnarray}
And according to \cite{Err}
\begin{equation}
  W_2 = \hat W_2 + \frac{85}{2304\pi^3}\zeta_3.
\end{equation}

\end{document}